\documentclass[epj]{svjour}
\usepackage{graphics}
\usepackage{times}
\begin{document}
\title{
Dynamical moments of inertia associated with wobbling motion 
in the triaxial superdeformed nucleus
}
\author{Masayuki Matsuzaki\inst{1}\thanks{Email address: matsuza@fukuoka-edu.ac.jp}
\and Yoshifumi R. Shimizu\inst{2}
\and Kenichi Matsuyanagi\inst{3}
}                     
%
%
\institute{Department of Physics, Fukuoka University of Education, Munakata, Fukuoka 811-4192, Japan
\and Department of Physics, Graduate School of Sciences, Kyushu University, Fukuoka 812-8581, Japan
\and Department of Physics, Graduate School of Science, Kyoto University, Kyoto 606-8502, Japan}
\date{Received: 17 October 2002}
\abstract{
The three moments of inertia associated with the wobbling mode built on the 
triaxial superdeformed states in Lu--Hf region are investigated 
by means of the cranked shell model plus random-phase 
approximation to the configurations with aligned quasiparticle(s). 
The result indicates that it is crucial to take into account the direct 
contribution to the moments of inertia from the aligned quasiparticle(s) 
so as to realize $\mathcal{J}_x > \mathcal{J}_y$ in positive-$\gamma$ shapes. 
} 
\authorrunning{Matsuzaki, Shimizu, and Matsuyanagi}
\titlerunning{Dynamical moments of inertia ...}
\maketitle
 The wobbling motion is a decisive evidence of stable triaxial deformations 
in rapidly rotating nuclei as discussed by Bohr and Mottelson about thirty years 
ago. 
Naming the main rotational axis the $x$ axis, its excitation energy is given by 
\[
\hbar\omega_\mathrm{wob}=\hbar\omega_\mathrm{rot}
\sqrt{
\frac{(\mathcal{J}_x-\mathcal{J}_y)(\mathcal{J}_x-\mathcal{J}_z)}
     {\mathcal{J}_y\mathcal{J}_z}} ,
\]
where $\omega_\mathrm{rot}$ is the rotational frequency of the main rotation 
about the $x$ axis and $\mathcal{J}$s are the three moments of inertia. 
Although there had been no definite experimental information about it so far, 
the first 
firm evidence in the triaxial superdeformed (TSD) states in $^{163}$Lu was 
reported last year\cite{lu1}. Whereas the expression above requires 
$\mathcal{J}_x>\mathcal{J}_y$ and $\mathcal{J}_z$, the observed TSD band is thought of 
as $\gamma>0$, which leads to $\mathcal{J}_x<\mathcal{J}_y$ in the irrotational 
moments of inertia,
\[
\mathcal{J}_k^\mathrm{irr}=\frac{4}{3}\mathcal{J}_0
\sin^2{(\gamma+\frac{2}{3}\pi k)} ,
\]
with $k=$ 1 -- 3 denoting the $x$, $y$ and $z$ principal axes, 
that are believed to describe realistic nuclei well. 
\begin{figure}[htbp]
\resizebox{0.4\textwidth}{!}{%
  \includegraphics{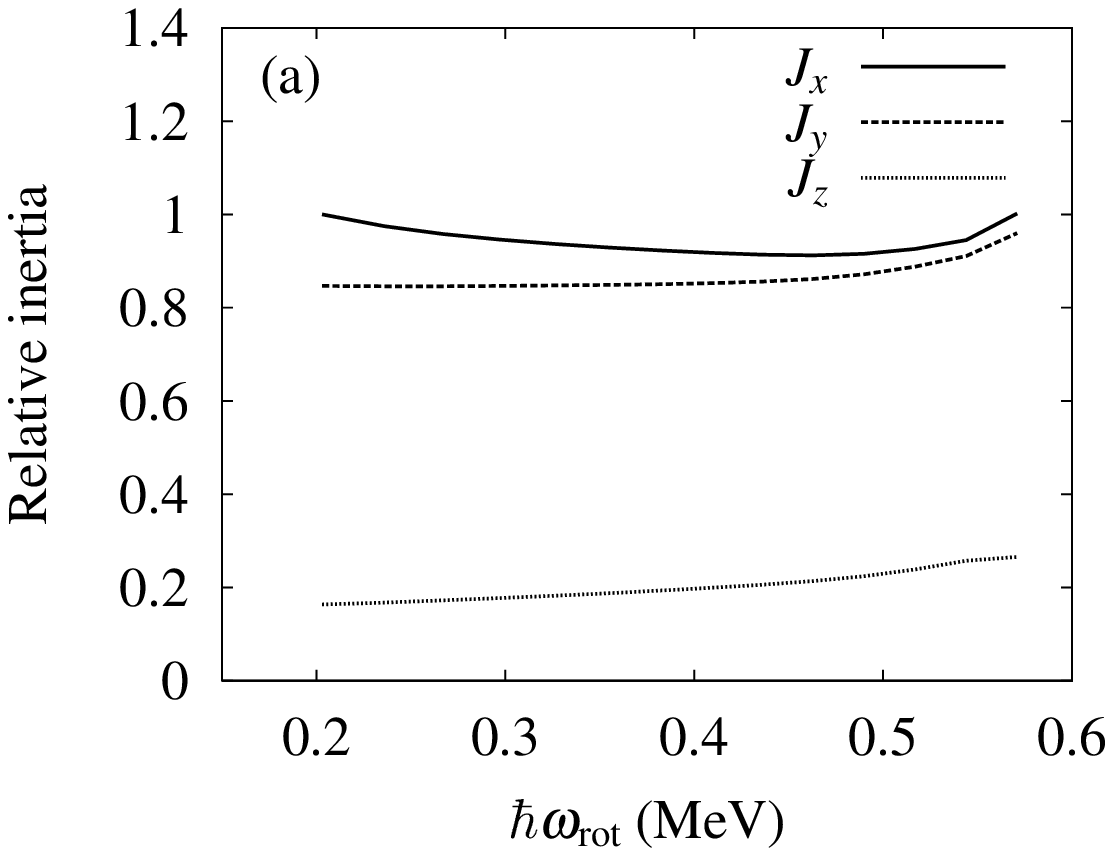}
}
\resizebox{0.4\textwidth}{!}{%
  \includegraphics{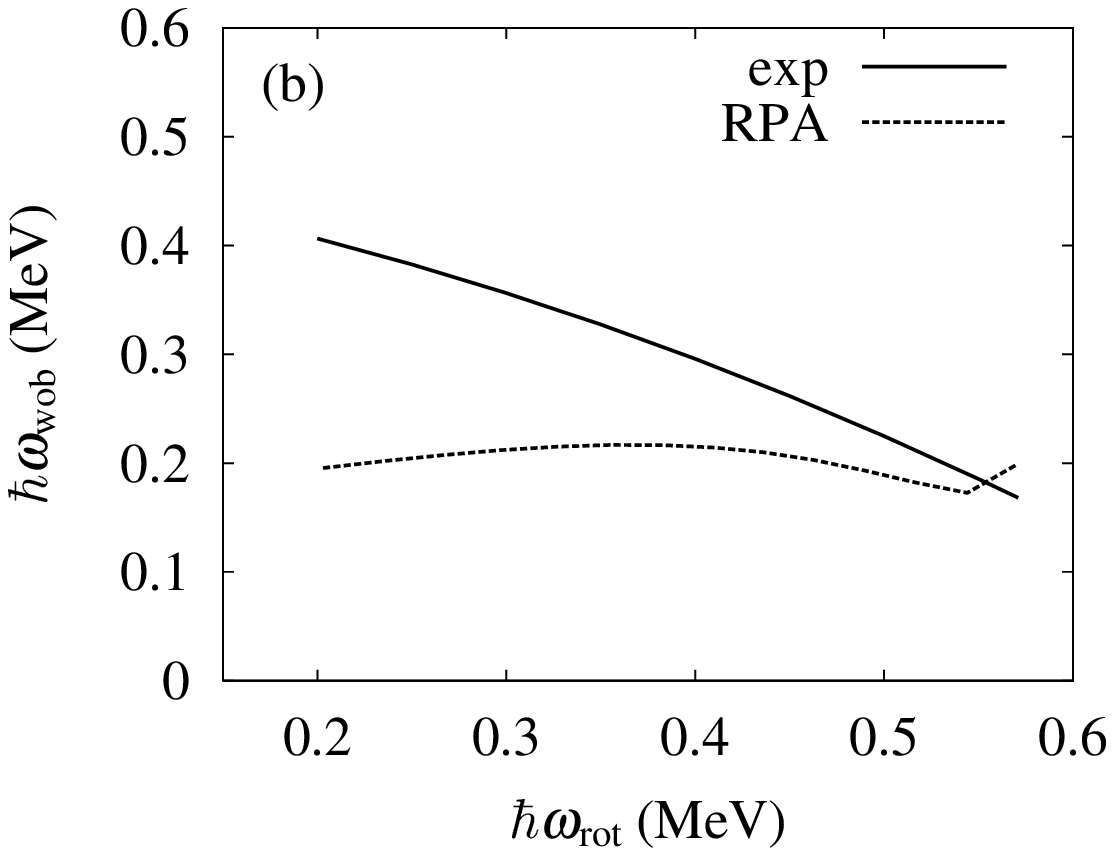}
}
\caption{(a) Calculated moments of inertia, and (b) experimental and calculated 
excitation energies of the wobbling mode in $^{163}$Lu. 
Note that the proton $BC$ crossing occurs at around 
$\hbar\omega_\mathrm{rot}\geq$ 0.55 MeV in the calculation. (Taken from Ref.\cite{msm}.)}
\label{fig:1}
\end{figure}

 Aiming at solving this puzzle, we performed a random-phase approximation (RPA) 
calculation in the rotating 
frame\cite{mj,ma,sm,smm} to study the dynamical moments of inertia,
$\mathcal{J}_y^\mathrm{(eff)}(\omega_\mathrm{wob})$ and 
$\mathcal{J}_z^\mathrm{(eff)}(\omega_\mathrm{wob})$,
determined simultaneously with the excitation energy $\hbar\omega_\mathrm{wob}$ 
by the dispersion equation, 
\[
\resizebox{0.45\textwidth}{!}{$\displaystyle
(\hbar\omega_\mathrm{wob})^2=(\hbar\omega_\mathrm{rot})^2
\frac{(\mathcal{J}_x-\mathcal{J}_y^\mathrm{(eff)}(\omega_\mathrm{wob}))
      (\mathcal{J}_x-\mathcal{J}_z^\mathrm{(eff)}(\omega_\mathrm{wob}))}
     {\mathcal{J}_y^\mathrm{(eff)}(\omega_\mathrm{wob})
      \mathcal{J}_z^\mathrm{(eff)}(\omega_\mathrm{wob})}$} ,
\]
which is derived from the coupled RPA equation of motion.
We have developed a computer code for the RPA to excitation modes built on 
configurations with arbitrary number of aligned quasiparticles (QPs). 
In this talk, we first study the dynamical moments of inertia, 
associated with the wobbling 
motion, of the whole system including the odd $\pi i_{13/2}$ quasiparticle. 
We did not perform full selfconsistent calculations
for given force strengths, but fixed them 
in such a way to guarantee the decoupling of the Nambu-Goldstone modes
for a given mean field.  Thus there is no ambiguity about the choice
of force strengths for the RPA calculations.
Our calculation\cite{msm} adopting $\epsilon_2=$ 0.43 and 
$\gamma=20^\circ$~\footnote{
This shape lead to transition quadrupole moments 
$Q_\mathrm{t}=$ 10.9 -- 11.3 $e$b 
for $\hbar\omega_\mathrm{rot}=$ 0.20 -- 0.57 MeV in accordance with the data, 
$Q_\mathrm{t}=10.7\pm0.7~e$b \cite{lu0}.
But it was reported by P. Fallon in this conference that a new measurement indicates 
that $Q_\mathrm{t}$ may be somewhat smaller.} 
proved that the contribution from the alignment of 
the intruder $\pi i_{13/2}$ quasiparticle, 
$\Delta \mathcal{J}_x=i_\mathrm{QP}/\omega_\mathrm{rot}$, 
superimposed on an irrotational-like inertia of the 0QP part, 
made the total $\mathcal{J}_x$ larger than $\mathcal{J}_y$ as shown in 
Fig.\ref{fig:1}(a). 

 Another virtue of the present framework is that the $\omega_\mathrm{rot}$ 
dependence of $\mathcal{J}$s is introduced automatically even when the 
mean field parameters are fixed. In particular, the decrease of 
$\mathcal{J}_x-\mathcal{J}_y$ due to a near constancy of the aligned angular 
momentum of the intruder $\pi i_{13/2}$ quasiparticle, $i_\mathrm{QP}$, 
leads to the decrease of 
$\omega_\mathrm{wob}$ as in Fig.\ref{fig:1}(b), whereas 
$\omega_\mathrm{rot}$-independent 
$\mathcal{J}$s lead to $\omega_\mathrm{wob}\propto\omega_\mathrm{rot}$. 

\begin{figure}
\resizebox{0.4\textwidth}{!}{%
  \includegraphics{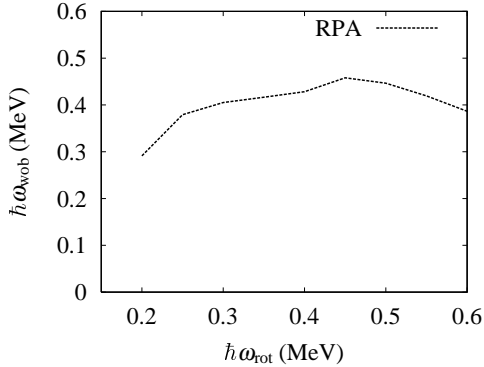}
}
\caption{Calculated excitation energy of the wobbling mode in $^{168}$Hf.}
\label{fig:2}
\end{figure}

 Next, in order to demonstrate that this mechanism applies also to 
2QP states in even-even nuclei, we present the result for $^{168}$Hf 
although the character of the excited TSD bands has not been clarified 
experimentally\cite{hf}. 
Figure \ref{fig:2} shows the result.
Thanks to one more aligned QP in comparison with the odd-$A$ Lu, 
the calculated $\omega_\mathrm{wob}$ is larger. 
In addition, the $(\nu j_{15/2})^2$ 
alignment occurs at around $\hbar\omega_\mathrm{rot}$ = 0.45 MeV, 
this makes $\omega_\mathrm{wob}$ even larger.

\begin{figure}
\resizebox{0.4\textwidth}{!}{%
  \includegraphics{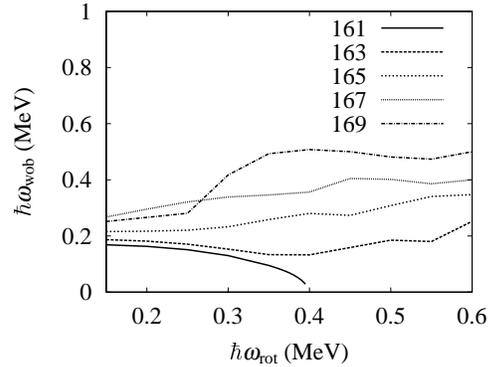}
}
\caption{Calculated excitation energies of the wobbling modes in $^{161-169}$Lu.} 
\label{fig:3}
\end{figure}

 Finally, we briefly mention the $N$-dependence since some new experimental information 
was presented in this conference. A preliminary calculation 
indicates that the $(\nu j_{15/2})^2$ aligns at lower $\omega_\mathrm{rot}$ 
in heavier isotopes, and consequently $\omega_\mathrm{wob}$ becomes larger in heavier 
isotopes as shown in Fig.\ref{fig:3}. 
Here, Nilsson model parameters slightly different from those in Fig.\ref{fig:1} are used.
Details will be discussed in Ref.\cite{msm2}.

 To summarize, we have performed the RPA calculation in the rotating 
frame to the triaxial superdeformed odd-$A$ nucleus $^{163}$Lu, in which 
the wobbling motion was observed for the first time, and discussed the 
physical conditions for its appearance. 
We have confirmed that the proton $i_{13/2}$ alignment that makes 
$\mathcal{J}_x>\mathcal{J}_y^\mathrm{(eff)}$ is 
indispensable for the appearance of the wobbling mode in this nucleus with a 
positive-$\gamma$ shape. 
This mechanism has been shown to apply also to even-even nuclei. 
A qualitative prediction for the $N$-dependence in odd-$A$ Lu isotopes has been 
briefly mentioned.


\begin{thebibliography}{99}
\bibitem{lu1}S. W. {\O}deg{\aa}rd {\it et al.}, Phys. Rev. Lett. \textbf{86}, (2001) 5866.
\bibitem{mj}I. N. Mikhailov and D. Janssen, Phys. Lett. \textbf{72B}, (1978) 303.
\bibitem{ma}E. R. Marshalek, Nucl. Phys. \textbf{A331}, (1979) 429.
\bibitem{sm}Y. R. Shimizu and K. Matsuyanagi, Prog. Theor. Phys. \textbf{72}, (1984) 799.
\bibitem{smm}Y. R. Shimizu and M. Matsuzaki, Nucl. Phys. \textbf{A588}, (1995) 559.
\bibitem{msm}M. Matsuzaki, Y. R. Shimizu, and K. Matsuyanagi, Phys. Rev. \textbf{C65}, 
(2002) 041303(R).
\bibitem{lu0}W. Schmitz {\it et al.}, Phys. Lett. \textbf{B303}, (1993) 230.
\bibitem{hf}H. Amro {\it et al.}, Phys. Lett. \textbf{B506}, (2001) 39.
\bibitem{msm2}M. Matsuzaki, Y. R. Shimizu, and K. Matsuyanagi, Phys. Rev. \textbf{C69}, 
(2004) 034325. 
\end{thebibliography}
\end{document}